\documentclass[prb,aps,amssymb,twocolumn,showpacs,amsmath]{revtex4}

\usepackage{graphicx}

\begin{document}

\title{Faraday Rotation Spectroscopy of Quantum-Dot Quantum Wells}

\author{Florian Meier}
\author{David D. Awschalom}
\affiliation{Center for Spintronics and Quantum Computation,
University of California, Santa Barbara, California 93106, USA }
\date{\today}

\begin{abstract}
Time-resolved Faraday rotation studies of CdS/CdSe/CdS quantum-dot
quantum wells have recently shown that the Faraday rotation angle
exhibits several well-defined resonances as a function of probe
energy close to the absorption edge. Here, we calculate the
Faraday rotation angle from the eigenstates of the quantum-dot
quantum well obtained with ${\bf k}\cdot {\bf p}$ theory. We show
that the large number of narrow resonances with comparable
spectral weight observed in experiment is not reproduced by the
level scheme of a quantum-dot quantum well with perfect spherical
symmetry. A simple model for broken spherical symmetry yields
results in better qualitative agreement with experiment.
\end{abstract}


\maketitle

\section{Introduction}

Semiconductor heterostructures with a size of several nanometers,
such as core-shell quantum dots, have been widely studied in the
past years. The combination of several semiconducting materials in
a single nanocrystal provides additional degrees of freedom which
allow one to tailor the properties of nanocrystals to a certain
extent.~\cite{alivisatos:96b,alivisatos:96} Capping layers of a
high-bandgap material on a low-bandgap quantum dot (QD) passivate
the surface and increase the optical
gain.\cite{hines:96,dabbousi:97} Quantum-dot quantum wells
(QDQWs), with several layers of a low-bandgap material sandwiched
between a high-bandgap core and cap are not only of interest from
a fundamental point of
view,\cite{kortan:90,mews:94,mews:96,tian:96,little:01,battaglia:03}
but could also be scaled up to functional heterostructures in a
single nanometer-size object.

Although QDQWs have been studied for several years, a detailed investigation of the quantum size spectrum using
photoluminescence (PL) or absorption spectroscopy is challenging because of inhomogeneous broadening. PL
typically shows a single broad peak, while the absorption spectrum exhibits a staircase-like spectrum without
distinct resonances. Individual exciton transitions have been observed using techniques such as hole burning
where a homogeneous subset of QDQWs is probed selectively.~\cite{mews:96} Time-resolved Faraday rotation (TRFR),
a well-established technique to investigate the spin dynamics in nanocrystals,~\cite{gupta:99,gupta:02} has
recently been used to characterize colloidal CdS/CdSe/CdS QDQWs with hexagonal wurtzite crystal
structure.~\cite{berezovsky:04} The TRFR signal amplitude was found to depend sensitively on probe energy and to
exhibit three resonances with linewidths as small as $10-20 \, {\rm meV}$ within an energy window of $0.2 \,{\rm
eV}$ around the absorption edge. Hence, TRFR does not only provide information on the spin dynamics, but also is
a sensitive spectroscopic technique which allows one to identify individual exciton transitions in QDQWs.

Here, we develop a microscopic theory for the TRFR signal amplitude as a function of probe energy for QDQWs.
From the eigenstates calculated using a two- and four-band ${\bf k}\cdot {\bf p}$ description for the conduction
and valence band states, respectively, we determine the dynamic dielectric response functions for $\sigma^\pm$
circularly polarized light and the amplitude of the TRFR signal, $\theta_F(E)$. While several narrow resonances
in $\theta_F(E)$ are predicted and the resonance energies are well reproduced by ${\bf k}\cdot {\bf p}$ theory,
our calculations show that the spectral weight of the resonances detected experimentally~\cite{berezovsky:04} is
not reproduced by the level scheme of a spherical QDQW (Sec.~\ref{sec:sphere}). We discuss a simple model in
which deformation of the QDQW leads to mixing of valence band multiplets, such that spectral weight is
re-distributed between the different dipole-allowed exciton transitions. Our calculations show that this model
yields better qualitative agreement with experimental data and may provide a possible explanation for the
well-defined resonances in the TRFR signal and the featureless increase in the absorption signal
(Sec.~\ref{sec:mixing}). In Sec.~\ref{sec:disc}, we summarize our results.

\section{Spherical QDQWs}
\label{sec:sphere}

\subsection{Energy level scheme}
\label{sec:spectrum}

\begin{figure}
\centerline{\mbox{\includegraphics[width=8.5cm]{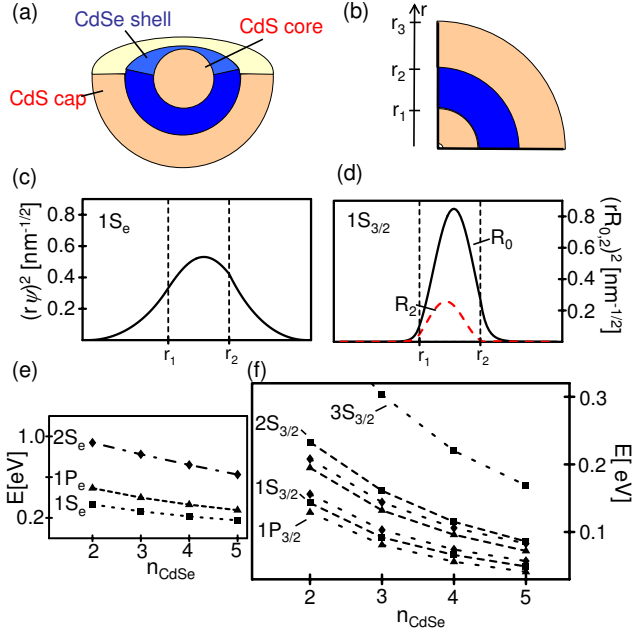}}}
\caption{(color online). (a) Schematic representation of the
CdS/CdSe/CdS QDQW of Ref.~\onlinecite{berezovsky:04}. (b) The
radius of the central CdS core is denoted by $r_1$, the width of
the CdSe QW by $r_2-r_1$, and the width of the CdS cap by
$r_3-r_2$. Numerical values are given in the text. (c) Radial
probability distribution of the conduction band ground state
$1S_e$. Because of the small conduction band mass, the state is
not well localized in the QW (indicated by dashed vertical lines).
(d) Radial probability distribution of the $1S_{3/2}$ valence band
state. The two components of the wave function, $R_0$ and $R_2$,
are shown in solid and dashed lines, respectively. Because the
valence band mass is large compared to the conduction band mass,
the valence band states are more strongly localized in the QW. (e)
Energies of the $1S_e$, $2S_e$, and $1P_e$ conduction band levels
as a function of QW width. (f) Energies of the lowest three
valence band levels with $S_{3/2}$-symmetry (squares) as a
function of QW width. The energies of the states $1P_{3/2}$,
$1D_{5/2}$, $1P_{5/2}$, and $1D_{7/2}$ are also shown (from
bottom). Labels for $1D_{5/2}$, $1P_{5/2}$, and $1D_{7/2}$ are
omitted for clarity. }\label{Fig1}
\end{figure}

We first consider CdS/CdSe/CdS QDQWs with perfect spherical
symmetry as shown in Figs.~\ref{Fig1}(a), (b). The CdS/CdSe/CdS
QDQWs have hexagonal wurtzite crystal structure. For the QDQWs in
Ref.~\onlinecite{berezovsky:04}, the radius of the CdS core and
the width of the CdS cap are  $r_1 \simeq 1.7\,{\rm nm}$ and
$r_3-r_2\simeq 1.6\,{\rm nm}$, respectively. The width of the CdSe
quantum well (QW), $r_2-r_1 = n_{\rm CdSe} a_{\rm CdSe}$, is
estimated from the number of CdSe monolayers, $n_{\rm CdSe}$, and
the monolayer thickness in bulk CdSe, $a_{\rm CdSe} \simeq 0.43\,
{\rm nm}$.~\cite{li:04} Because the focus of this paper is the
TRFR signal amplitude as a function of probe energy, we restrict
ourselves to the simplest realistic description of the QDQW. The
energy levels are calculated from ${\bf k}\cdot {\bf p}$
theory,~\cite{jaskolski:98,pokatilov:01} with a two-band
Hamiltonian for the conduction band states and a four-band
Luttinger Hamiltonian for the heavy and light hole valence band
states.

In the two-band approximation, the conduction band states are
determined by
\begin{equation} \hat{H}_{\rm cb} = \hat{\bf k}\frac{1}{2 m(r)}\hat{\bf k} + V_c(r), \label{eq:cb-ham}
\end{equation}
where the band mass is given by the value of bulk CdSe and CdS,
respectively,~\cite{landolt}
\begin{equation}
m(r) = \left\{ \begin{array}{cc}  m_{\rm CdSe}=0.11 \, m_0, &  \, r_1 < r < r_2, \\
m_{\rm CdS} = 0.15 \, m_0, & \, r < r_1 \, \, {\rm or} \, r> r_2.\end{array} \right. \label{eq:cbmass}
\end{equation}
$m_0$ denotes the free electron mass and $\hat{\bf k}$ the
momentum operator for the envelope wave function. The potential
$V_c(r)$ in $\hat{H}_{\rm cb}$ represents the offset  of the CdS
conduction band relative to CdSe, $0.32\,~{\rm eV}$.~\cite{wei:00}
Vanishing boundary conditions are imposed at $r_3$. In Fig.~1(e),
the $1S_e$, $2S_e$, and $1P_e$ energy levels are displayed as a
function of QW width. Figure~\ref{Fig1}(c) shows the radial
probability distribution for the conduction band ground state,
$1S_e$, for $n_{\rm CdSe}=3$. Because of the small conduction band
mass, the state is not well localized in the QW. The lattice
mismatch at the CdS/CdSe interface ($4$\%) is expected to also
modify the electronic band structure,~\cite{peng:97} an effect
that is neglected in our calculations. While the detailed band
structure close to the interface is not fully understood, a likely
scenario is that the lattice constant varies gradually, giving
rise to band realignment and a gradual change in the radial
potential $V_c(r)$ rather than the step-like behavior considered
here.

In the spherical approximation, heavy and light hole valence band
states are determined by the Luttinger Hamiltonian
\begin{equation}
\hat{H}_{\rm vb} = [\gamma_1(r) + \frac{5}{2}\gamma
(r)]\frac{\hat{\bf k}^2}{2 m_0} - \frac{\gamma(r)}{m_0} (\hat{\bf
k} \cdot \hat{\bf J} )^2 + V_v(r) \label{eq:vb-ham}
\end{equation}
for $r\neq r_{1,2}$. $\hat{\bf J}$ denotes the spin operator of
the $J=3/2$ valence band multiplet. States of the $J=1/2$
split-off band are offset in energy by $0.4 \, {\rm eV}$ and are
neglected in the following. $\gamma_1(r)$ and $\gamma (r)$ are the
Luttinger parameters of CdSe (CdS) for $r$ inside (outside) the
QW,~\cite{richard:96,landolt,rem1}
\begin{equation}
\gamma_1(r) = \left\{ \begin{array}{cc}  \gamma_{1,{\rm CdSe}}=1.67, &  \, r_1 < r < r_2; \\
\gamma_{1,{\rm CdS}}= 1.09, & \, r < r_1 \, \, {\rm or} \, r>
r_2,\end{array} \right. \label{eq:ll1param}
\end{equation}
and
\begin{equation}
\gamma(r) = \left\{ \begin{array}{cc}  \gamma_{\rm CdSe}=0.56, &  \, r_1 < r < r_2; \\
\gamma_{\rm CdS}= 0.34, & \, r < r_1 \, \, {\rm or} \, r>
r_2.\end{array} \right. \label{eq:llparam}
\end{equation}
Because $\hat{\bf k}$ does not commute with $\gamma_1(r)$ and
$\gamma(r)$, operator ordering in Eq.~(\ref{eq:vb-ham}) is
important for $r=r_{1,2}$ (see Ref.~\onlinecite{pokatilov:01} and
references therein) and will be specified by the boundary
conditions at the interfaces in Eq.~(\ref{eq:vb-bc}) below.
$V_v(r)$ represents the offset of $0.42\,~{\rm eV}$ of the CdS
valence band edge relative to CdSe.~\cite{wei:00}

The eigenstates of $\hat{H}_{\rm vb}$ have been explicitly
calculated for both QDs~\cite{xia:89,efros:92} and
QDQWs.~\cite{jaskolski:98,pokatilov:01} While the orbital angular
momentum $\widehat{\bf L}$ of the envelope wave function is no
longer a good quantum number, the sum of orbital and spin angular
momentum, $\widehat{\bf F} = \widehat{\bf L} + \widehat{\bf J}$,
commutes with the Hamiltonian, $[\hat{H}_{\rm vb},\widehat{\bf
F}]=0$. Eigenstates of $\hat{H}_{\rm vb}$ are labelled according
to the quantum numbers $F$, $F_z$, and the smallest angular
momentum component $L$ of the envelope wave function. The
eigenstates~\cite{xia:89,efros:92}
\begin{equation} |n L_F;F_z\rangle = R_L (r) |L,{\textstyle
\frac{3}{2}},F,F_z\rangle + R_{L+2}  (r) |L+2,{\textstyle
\frac{3}{2}},F,F_z\rangle \label{eq:states}
\end{equation}
are superpositions of envelope functions with angular momentum $L$
and $L+2$. For $r<r_1$, $r_1<r<r_2$, and $r_2<r$, the radial wave
functions $R_L(r)$ and $R_{L+2}(r)$ are solutions of the
differential equations~\cite{xia:89}
\begin{widetext}
\begin{eqnarray}
&& \left( \begin{array}{cc} -\frac{\hbar^2}{2 m_0} (\gamma_1 + c_1
\gamma)\left( \partial_r^2 +\frac{2}{r} \partial_r -
\frac{L(L+1)}{r^2}\right)  + V_{v}(r) & c_2 \gamma
\frac{\hbar^2}{2 m_0} \left( \partial_r^2 +\frac{2L+5}{r}
\partial_r +
\frac{(L+1)(L+3)}{r^2}\right) \\
c_2 \gamma  \frac{\hbar^2}{2 m_0} \left( \partial_r^2
-\frac{2L+1}{r}
\partial_r + \frac{L(L+2)}{r^2}\right) &
-\frac{\hbar^2}{2 m_0} (\gamma_1 - c_1 \gamma)\left( \partial_r^2
+\frac{2}{r} \partial_r - \frac{(L+2)(L+3)}{r^2}\right)  +
V_{v}(r)
\end{array} \right) \left(\begin{array}{c} R_L \\
R_{L+2}
\end{array} \right) \nonumber \\ && \hspace*{12cm}= E \left(\begin{array}{c} R_L \\ R_{L+2}
\end{array} \right).  \label{eq:rad-deq}
\end{eqnarray}
\end{widetext}
The full Hamiltonian with the correct operator ordering at
$r=r_{1,2}$ is given, e.g.,  in Ref.~\onlinecite{pokatilov:01} and
is omitted here for brevity. The dimensionless constants $c_{1,2}$
depend on $F$ and $L$ and can be read off from the values
$C_{1,2}$ in Table I of Ref.~\onlinecite{xia:89} with the
correspondence $c_{1,2} = 2 C_{1,2}/\mu$. Equation
(\ref{eq:rad-deq}) is solved with a piecewise ansatz in the
spherical Bessel functions $j_L$ and $n_L$. The boundary
conditions for the derivatives require that
\begin{eqnarray}
&& \left( \begin{array}{cc}  \gamma_1 \partial_r + c_1 \gamma
\left(
\partial_r +\frac{3}{2r} \right)  & - c_2
\gamma \left( \partial_r +\frac{L+3}{r}\right) \\
-c_2 \gamma  \left( \partial_r -\frac{L}{r} \right) & \gamma_1
\partial_r - c_1 \gamma \left( \partial_r +\frac{3}{2r} \right)
\end{array} \right) \nonumber   \\ && \hspace*{5cm} \times
\left(\begin{array}{c} R_L \\ R_{L+2}
\end{array} \right) \label{eq:vb-bc}
\end{eqnarray}
is continuous at $r_{1,2}$.~\cite{pokatilov:01}

We have calculated the lowest conduction and valence band energy
levels for the QDQWs of Ref.~\onlinecite{berezovsky:04}. The wave
functions of the conduction band ground state, $1S_e$, and the
components $R_{0,2}$ of $1S_{3/2}$ are shown in
Figs.~\ref{Fig1}(c) and (d), respectively. The lowest valence band
states are (in order of increasing energy) $1P_{3/2}$, $1S_{3/2}$,
$1D_{5/2}$, and $1P_{5/2}$ [Fig.~~\ref{Fig1}(f)]. Similarly to
CdS/HgS/CdS
QDQWs,~\cite{schooss:94,jaskolski:98,pokatilov:01,bryant:03} the
envelope function of the valence band ground state is $p$-type,
implying a dark exciton ground state. While the conduction band
states are strongly delocalized over core, QW, and cap, the
valence band states are much better localized in the QW because of
the larger valence band mass.

The electron-hole Coulomb attraction is calculated in first order
perturbation theory because the QDQWs are in the
strong-confinement regime. We neglect the small difference in
dielectric constants for CdS and CdSe and set
$\epsilon=\epsilon_{\rm CdSe} = \epsilon_{\rm CdS} = 9$. For an
electron in the conduction band ground state, $1S_e$, the exciton
binding energy $U$ is
\begin{eqnarray} U &=& - \frac{e^2}{4 \pi
\epsilon  \epsilon_0} \int_0^{r_3} d r_e \, dr_h \, \frac{r_e^2
r_h^2}{\max (r_e,r_h)} \psi_{1S_e}(r_e)^2 \label{eq:coul-int} \\
&&  \hspace*{2cm} \times [R_L(r_h)^2 + R_{L+2}(r_h)^2], \nonumber
\end{eqnarray}
with $\psi_{1S_e}(r_e)$ the radial wave function of $1S_e$ and
$R_{L,L+2}$ the radial components of the hole state. For both the
$1P_{3/2}$ and the $1S_{3/2}$ multiplet, the Coulomb integral is
well approximated by $U = -e^2/4 \pi \epsilon \epsilon_0
[(r_1+r_2)/2]$. The calculation of the dynamic dielectric response
functions in Sec.~\ref{sec:diel} involves virtual transitions from
one-exciton to bi-exciton states, such that the bi-exciton shift
must be evaluated. For a bi-exciton with two electrons in $1S_e$
and two holes in $1S_{3/2}$, the bi-exciton shift is of order $5\,
{\rm meV}$, sufficiently small that it can be neglected in the
following. Similarly, the electron-hole exchange interaction is
neglected because its characteristic energy scale is of order
$1\,{\rm meV}$.~\cite{norris:96}

The lattice anisotropy of the wurtzite crystal structure is taken
into account by the anisotropy Hamiltonian~\cite{efros:92}
\begin{equation}
\hat{H}_{\rm an}=\Delta \left[ (3/2)^2 -\hat{J}_z^2 \right].
\label{eq:anisotropy}
\end{equation}
Because the hole wave functions are localized predominantly in the
CdSe QW, we approximate $\Delta = 25\,{\rm meV}$ by the CdSe bulk
value.~\cite{efros:92,efros:98} $\hat{H}_{\rm an}$ lifts the
degeneracy of an $L_F$-multiplet and splits it into $(2F+1)/2$
energy doublets. In particular, for $S_{3/2}$ multiplets, the
energy shifts induced by $\hat{H}_{\rm an}$ are~\cite{efros:98}
\begin{subequations}
\label{eq:anshift}
\begin{eqnarray}
\Delta_{n,3/2} &=& \langle nS_{3/2};3/2 |\hat{H}_{\rm
an}|nS_{3/2};3/2 \rangle \label{eq:anshift1} \\ &=& \Delta
\frac{4}{5} \int dr \, r^2 R_2^2,\nonumber  \\
\Delta_{n,1/2} &=& \langle nS_{3/2};1/2 |\hat{H}_{\rm
an}|nS_{3/2};1/2 \rangle \label{eq:anshift2} \\ &=& \Delta \Bigl[
\int dr \, r^2 R_0^2+\frac{1}{5} \int dr \, r^2 R_2^2 \Bigr]
\nonumber
\end{eqnarray}
\end{subequations}
for $|F_z| = 3/2$ and $|F_z|=1/2$, respectively.

\subsection{Faraday rotation angle}
\label{sec:diel}

We next calculate the amplitude of the Faraday rotation (FR) angle
as a function of probe energy, $\theta_F(E)$, taking into account
the single-particle levels, Coulomb interaction, and crystal
anisotropy as described in Sec.~\ref{sec:spectrum}. In the
pump-probe scheme of Ref.~\onlinecite{berezovsky:04}, a $\sigma^-$
circularly polarized pump pulse with energy large compared to the
absorption edge excites spin-polarized excitons. While the
conduction band electron typically retains its spin polarization
on relaxation to the conduction band ground state, $1S_e$, the
hole spin is believed to randomize quickly during relaxation to
$1P_{3/2}$. The net spin polarization along the pump direction,
which results from conduction band electrons in the spin state
$|s_z = \uparrow\rangle$, is experimentally detected with TRFR.
For probe energy $E$, $\theta_F(E)$ is proportional to the
difference of the dynamic dielectric response functions for
$\sigma^\pm$ circularly polarized light, $\epsilon_+(E/\hbar) -
\epsilon_-(E/\hbar)$, which are determined by the optical dipole
transition matrix
elements,~\cite{hugonnard:94,linder:98,sham:99,meier:04}
\begin{eqnarray}
\theta_F(E)& = &C E \sum_{\sigma =\pm 1; |XX \rangle} \sigma
\left| \langle XX |\hat{p}_x + \sigma  i \hat{p}_y |X_{\rm in}
\rangle \right|^2  \nonumber \\ && \hspace*{1cm} \times
\frac{E-(E_{XX}-E_{X_{\rm in}})}{(E_{XX}-E_{X_{\rm in}})^2 +
\gamma_{XX}^2}. \label{eq:frot1}
\end{eqnarray}
$|X_{\rm in}\rangle$ denotes the initial $1S_e-1P_{3/2}$ exciton
prepared by the pump pulse with an $|s_z =
\uparrow\rangle$-electron in $1S_e$. The sum extends over all
bi-exciton states $|XX \rangle$ with energy $E_{XX}$, and
$\gamma_{XX}$ denotes the linewidth of the corresponding
bi-exciton transition. For simplicity, we assume that the
linewidth of all transitions is equal, $\gamma_{XX} = \Gamma$. The
constant $C$ depends on the sample size and refractive index.
$\theta_F(E)$ is finite because transitions to bi-exciton states
with both conduction band electrons in $1S_e$ are allowed only if
the electrons form a singlet state. Hence, the matrix elements in
Eq.~(\ref{eq:frot1}) can be expressed in terms of the transitions
from an arbitrary valence band state $|\Phi_v\rangle$ to the
unoccupied conduction band state $|1S_e;\downarrow\rangle$,
\begin{eqnarray}
\theta_F(E)& = &C E \sum_{\sigma =\pm 1; |\Phi_v\rangle} \sigma
\left| \langle 1S_e;\downarrow |\hat{p}_x + \sigma  i
\hat{p}_y |\Phi_v  \rangle \right|^2  \nonumber \\
&&  \hspace*{2cm} \times \frac{E-E_{X,v}}{(E-E_{X,v})^2 +
\Gamma^2}, \label{eq:frot2}
\end{eqnarray}
with $E_{X,v}$ the energy of the $1S_e$-$\Phi_v$ exciton. As
mentioned in Sec.~\ref{sec:spectrum}, the bi-exciton shift is
negligible. This expression depends only on the single-particle
levels in the conduction and valence band and can be evaluated
from the wave functions calculated above.

The dipole transition matrix elements are evaluated following
Ref.~\onlinecite{efros:92}. The overlap integral of the envelope
wave functions is finite only for $s$-type valence band states,
$|\Phi_v\rangle=|nS_{3/2};F_z\rangle$. We first consider a QDQW
with crystal symmetry axis aligned along the laser direction.
Then,
\begin{subequations}
\label{eq:mel1}
\begin{eqnarray}
&& |\langle 1S_e;\downarrow |\hat{p}_x -  i \hat{p}_y
|nS_{3/2};F_z\rangle |^2   \label{eq:mel1a}
\\ && \hspace*{1cm} = \frac{2}{3} \left| \int dr r^2 \psi_{1S_e} R_0
\langle S |\hat{p}_z | Z\rangle \right|^2 \delta_{F_z,1/2}
\nonumber
\\ && \hspace*{1cm} \simeq  \frac{2}{3} (m_0 V)^2 \left| \int dr r^2 \psi_{1S_e} R_0
\right|^2 \delta_{F_z,1/2},  \nonumber \\
&&  |\langle 1S_e;\downarrow |\hat{p}_x +  i \hat{p}_y
|nS_{3/2};F_z\rangle |^2 \label{eq:mel1b} \\
&& \hspace*{1cm} = 2  \left| \int dr r^2 \psi_{1S_e} R_0 \langle S
|\hat{p}_z | Z\rangle \right|^2 \delta_{F_z,-3/2} \nonumber
\\ && \hspace*{1cm} \simeq 2 (m_0 V)^2 \left| \int dr r^2 \psi_{1S_e} R_0
\right|^2 \delta_{F_z,-3/2} \nonumber ,
\end{eqnarray}
\end{subequations}
while the matrix elements vanish for $F_z = 3/2, -1/2$. The Kane
interband matrix element $\langle S |\hat{p}_z | Z\rangle$ varies
spatially for the QDQW, but because the valence band states are
well localized in the CdSe QW, we approximate $|\langle S
|\hat{p}_z | Z\rangle| \simeq m_0 V$ by the CdSe interband matrix
element. The overlap integral in Eq.~(\ref{eq:mel1}), in the
following denoted by
\begin{equation}
I_n = \left| \int dr r^2 \psi_{1S_e} R_0 \right|^2,
\label{eq:radint}
\end{equation}
depends on the quantum number $n$ via $R_0$. For $1S_{3/2}$,
$2S_{3/2}$, and $3S_{3/2}$, the numerical values are given in
Table~\ref{tab1} for different QW widths.

\begin{table}[t!]
\begin{tabular}{c|c|c|c|c}
$n_{\rm CdSe}$ &  $3$ & $4$ & $5$ \\
\hline $I_1$ &  $0.60$ & $0.67$ & $0.72$ \\
\hline $I_2$ & $0.16$ & $0.11$ & $0.08$ \\
\hline $I_3$  & $0.09$ & $0.09$ & $0.10$ \\
\end{tabular}
\caption{Radial overlap integrals, Eq.~(\ref{eq:radint}), for
$1S_{3/2}$, $2S_{3/2}$, and $3S_{3/2}$ valence band states as a
function of the QW width, $n_{\rm CdSe}$.} \label{tab1}
\end{table}

With the transition matrix elements, $\theta_F(E)$ is readily
evaluated for $E$ close to the absorption edge. We define the
energy $E_{X,n} = E_{1S_e} + E_{nS_{3/2}} + E_g + U$ of the
$1S_e-nS_{3/2}$ exciton, where $E_g \simeq 1.75 \, {\rm eV}$ is
the bandgap of bulk CdSe at room temperature and $U$ the Coulomb
integral [Eq.~(\ref{eq:coul-int})]. Combining
Eqs.~(\ref{eq:frot2}) and (\ref{eq:mel1}),
\begin{eqnarray}
\theta_F (E) & \simeq & C^\prime E \sum_{n} I_n \Bigl[
\frac{E-(E_{X,n}+\Delta_{n,3/2})}{[E-(E_{X,n}+\Delta_{n,3/2})]^2 +
\Gamma^2}  \nonumber  \\ && \hspace*{0.5cm} - \frac{1}{3}
\frac{E-(E_{X,n}+\Delta_{n,1/2})}{[E-(E_{X,n}+\Delta_{n,1/2})]^2 +
\Gamma^2}\Bigr], \label{eq:frot3}
\end{eqnarray}
where $C^\prime =  2 C (m_0 V)^2$ and $E$ was assumed to be
sufficiently close to the absorption edge that transitions from
the split-off band can be neglected. The energy shifts
$\Delta_{n,3/2}$ and $\Delta_{n,1/2}$ are induced by the
structural anisotropy [Eq.~(\ref{eq:anshift})].

\begin{figure}
\centerline{\mbox{\includegraphics[width=8.5cm]{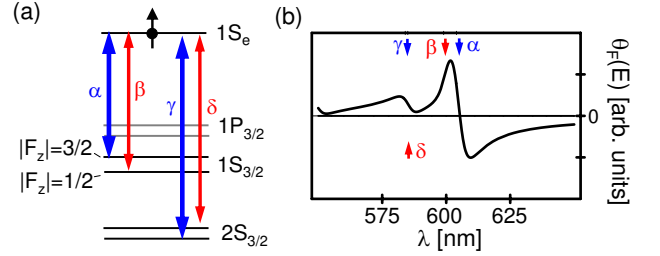}}} \caption{(color online). (a) Schematic
representation of the transitions which contribute to the FR angle, $\theta_F(E)$, in the spectral
representation Eq.~(\ref{eq:frot2}). The dipole transition matrix elements with $1S_e$ are finite only for
$S_{3/2}$-multiplets. (b) $\theta_F(E)$ calculated from Eq.~(\ref{eq:frot3}) for $n_{\rm CdSe}=3$ and $\Gamma=15
\, {\rm meV}$. $nS_{3/2}$ multiplets with $n\geq 4$ have been neglected. The transitions from (a) are indicated
by arrows. }\label{Fig2}
\end{figure}

For every valence band multiplet $nS_{3/2}$, $\theta_F(E)$ shows a double-resonance structure, with a main
resonance corresponding to transitions from $|F_z|=3/2$ [first term in the square bracket in
Eq.~(\ref{eq:frot3})] and a second resonance with $1/3$ smaller spectral weight that corresponds to transitions
from $|F_z|=1/2$ states and is shifted by $\Delta_{n,1/2}-\Delta_{n,3/2}$. The second resonance can only be
resolved if $\Gamma < \Delta$. For the colloidal QDQWs in Ref.~\onlinecite{berezovsky:04}, the inhomogeneous
line broadening is substantial, $\Gamma \gtrsim 15 \, {\rm meV}$, and the splitting of resonances caused by
crystal anisotropy cannot be resolved. The lowest transitions with finite matrix elements in the spectral
representation in Eq.~(\ref{eq:frot2}) are shown schematically in Fig.~\ref{Fig2}(a). Figure~\ref{Fig2}(b) shows
$\theta_F(E)$ close to the absorption edge for $n_{\rm CdSe} = 3$. While resonances corresponding to $1S_{3/2}$,
$2S_{3/2}$, and $3S_{3/2}$ can be resolved, the spectral weight for $1S_{3/2}$ is significantly larger than that
for $2S_{3/2}$ and $3S_{3/2}$. The behavior in Fig.~\ref{Fig2}(b) is clearly distinct from experimental
data~\cite{berezovsky:04} which exhibit three resonances with comparable spectral weight within $0.2\, {\rm eV}$
of the absorption edge.

We next calculate $\theta_F(E)$ for the case that the QDQW
symmetry axis is not aligned with the laser. By averaging over
different relative orientations, we also determine $\theta_F(E)$
for an ensemble of randomly oriented QDQWs. If the QDQW crystal
symmetry axis is tilted by $\psi$ relative to the laser
propagation direction, the electric field components of the pump
and probe laser pulses, given in the laboratory coordinate system
(indicated with a superscript $L$ in the following), are rotated
relative to the spin quantization axis of the QDQW. The spectral
representation for $\theta_F(E)$ is still given by
Eq.~(\ref{eq:frot2}), but the matrix elements must be modified to
$\langle 1S_e;\downarrow^L |\hat{p}_x^L + \sigma  i \hat{p}_y^L
|\Phi_v \rangle $ to account for the fact that the pump and probe
pulses are defined in the laboratory coordinate system $L$. To
evaluate the matrix elements, operators in the $L$-frame are
expressed in terms of the QDQW coordinate system, $|\!
\downarrow^L\rangle = i \sin(\psi/2) |\! \uparrow\rangle +
\cos(\psi/2) |\! \downarrow\rangle$, $\hat{p}_x^L = \cos (\psi)
\hat{p}_x + \sin (\psi) \hat{p}_z$, and $\hat{p}_y^L = \hat{p}_y$.
With the convention for the valence band basis functions in
Refs.~\onlinecite{luttinger:55,efros:98},
\begin{eqnarray}
&& \langle 1S_e; \downarrow^L | \hat{p}_x^L \pm  i \hat{p}_y^L
|nS_{3/2};F_z\rangle = \sqrt{I_n/2} \, m_0 V \label{eq:tilt-trel}\\
&& \times  \left\{\begin{array}{cc} \vspace*{0.2cm} -
i\sin(\psi/2) [\cos(\psi) \mp 1]
 , & \,\,\, F_z=3/2; \\
 -(2/\sqrt{3})  \sin(\psi/2) \sin(\psi) & \,\,\, F_z=1/2; \\
 \hspace*{0.5cm} \vspace*{0.2cm}
 + (i/\sqrt{3})
  \cos(\psi/2) [\cos(\psi) \mp 1], &  \\
 (2/\sqrt{3}) \cos(\psi/2) \sin(\psi)  & \,\,\, F_z=-1/2; \\
 \hspace*{0.5cm} \vspace*{0.2cm}
 - (i/\sqrt{3})
  \sin(\psi/2) [\cos(\psi) \pm 1], &  \\
 i\cos(\psi/2) [\cos(\psi) \pm 1]
 , & \,\,\, F_z=-3/2.
\end{array} \right. \nonumber
\end{eqnarray}

Inserting these transition matrix elements into the spectral
representation in Eq.~(\ref{eq:frot2}), we find that the
expression for $\theta_F(E)$ only acquires a pre-factor $\cos^2
(\psi/2) \, \cos (\psi)$,
\begin{eqnarray}
\theta_F (E) & \simeq & C^\prime E \cos^2 (\psi/2) \,  \cos (\psi)
 \label{eq:frot3-rand} \\ && \times
\sum_{n} I_n \Bigl[
\frac{E-(E_{X,n}+\Delta_{n,3/2})}{[E-(E_{X,n}+\Delta_{n,3/2})]^2 +
\Gamma^2}  \nonumber  \\ && \hspace*{0.5cm} - \frac{1}{3}
\frac{E-(E_{X,n}+\Delta_{n,1/2})}{[E-(E_{X,n}+\Delta_{n,1/2})]^2 +
\Gamma^2}\Bigr], \nonumber
\end{eqnarray}
but the relative weight of the individual terms is not altered. In
particular, Eq.~(\ref{eq:frot3-rand}) implies $\theta_F=0$ for
$\psi = \pi/2$. This can be understood in physical terms because
TRFR probes the spin polarization along the QDQW anisotropy axis,
while, for $\psi = \pi/2$, the pump pulse only generates spin
polarization perpendicular to the anisotropy axis.

Averaging Eq.~(\ref{eq:frot3-rand}) over an ensemble of randomly
oriented QDQWs is performed by integration over $\psi$. For random
QDQW orientation, Eq.~(\ref{eq:frot3}) remains valid with the
substitution $C^\prime \rightarrow 5 C^\prime/12$. While random
QDQW orientation reduces the total amplitude of $\theta_F(E)$, the
relative spectral weight of the individual contributions is not
altered and $\theta_F(E)$ is still as shown in Fig.~\ref{Fig2}(b).

\section{Broken spherical symmetry}
\label{sec:mixing}

The calculations in Sec.~\ref{sec:sphere} above show that, for a
spherical QDQW, $\theta_F(E)$ exhibits a pair of resonances for
every $nS_{3/2}-1S_e$ exciton transition. However, the spectral
weight of the resonances decreases rapidly with increasing $n$
(Table~\ref{tab1}), such that the large number of resonances with
comparable spectral weight observed
experimentally~\cite{berezovsky:04} is not correctly reproduced by
Eq.~(\ref{eq:frot3}). The experimental data imply that spectral
weight is redistributed from the $1S_{3/2}-1S_e$ exciton line to
other transitions.

We show next that broken spherical symmetry is a possible mechanism which accounts for the experimental FR data
by mixing of the $1S_{3/2}$ and $1P_{3/2}$ valence band multiplets. The admixture of $s$-type to $p$-type
multiplets redistributes the spectral weight and increases the number of resonances with comparable amplitude in
$\theta_F(E)$.  As will be shown below, the redistribution of spectral weight also explains the absence of
pronounced resonances in the absorption signal. Compared to spherical QDQWs, broken symmetry gives rise to a
larger energy splitting between the lowest valence band states with dominant $p$-type and $s$-type envelope wave
functions, consistent with the large Stokes shift between the PL peak and the absorption edge observed for
CdS/CdSe/CdS QDQWs.~\cite{battaglia:03,berezovsky:04}

\begin{figure}
\centerline{\mbox{\includegraphics[width=8.5cm]{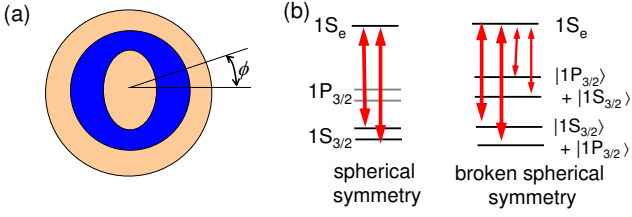}}}
\caption{(color online). (a) Cross section of a QDQW with broken
spherical symmetry. A deformation of the central CdS QD implies a
spatial variation of the QW width, which is modelled by
Eq.~(\ref{eq:symbreak}). (b) Schematic representation of the level
scheme for a spherical QDQW and a QDQW with broken spherical
symmetry. Broken spherical symmetry mixes the $1S_{3/2}$ and
$1P_{3/2}$ multiplets and renders transitions to $1S_e$ bright for
all eight states of the multiplet.}\label{Fig3}
\end{figure}

On a microscopic level, broken spherical symmetry would most
probably result from a spatial variation of the QW width [shown
schematically in Fig.~\ref{Fig3}(a)]. For QWs with $n_{\rm CdSe} =
2,3,4,5$, a monolayer variation in QW width translates into a
perturbation with a typical energy scale of order $0.1 \,{\rm eV}$
for the hole states. Conduction band states are less strongly
affected because they are not localized in the QW and the energy
level spacing is much larger. We do not attempt to describe
breaking of the spherical symmetry microscopically, but restrict
our discussion to a simple model in which the valence band
Hamiltonian includes a potential
\begin{equation}
\delta V ({\bf r}) = v_0 \sin \theta \left(1 + \cos \phi \right)
\label{eq:symbreak}
\end{equation}
which mixes $1S_{3/2}$ and $1P_{3/2}$. $\theta$ and $\phi$ denote
the azimuthal and polar angle of ${\bf r}$ relative to the lattice
symmetry axis, respectively. $v_0$ is a fit parameter.

The exact hole states with the perturbation
Eq.~(\ref{eq:symbreak}) can, at least in principle, be obtained by
diagonalization of $\delta V({\bf r})$ in the basis
$|nL_F;F_z\rangle$ of the spherical system. The only finite matrix
elements are $\langle n^\prime L^\prime_{F^\prime};F_z^\prime |
\delta V ({\bf r}) |n L_F;F_z\rangle \propto \delta
_{F_z,F_z^\prime}$ for $|L-L^\prime|$ an integer multiple of $2$
and $\langle n^\prime L^\prime_{F^\prime};F_z^\prime | \delta V
({\bf r})|n L_F;F_z\rangle \propto \delta _{F_z,F_z^\prime \pm 1}$
for $|L-L^\prime|=1$ or $|L-L^\prime|=3$. In order to keep the
subsequent discussion transparent, we restrict the diagonalization
of $\delta V ({\bf r})$ to the subspace spanned by the valence
band multiplets $1S_{3/2}$, $2S_{3/2}$,  $3S_{3/2}$, and
$1P_{3/2}$. These states are dominant for the optical response
close to the absorption edge and we expect this procedure to
qualitatively capture the essential features of broken spherical
symmetry.~\cite{rem2}

The off-diagonal matrix elements of $\delta V ({\bf r})$ between
the $1S_{3/2}$ and $1P_{3/2}$ multiplet are calculated by
expansion of the envelope wave functions in spherical harmonics.
We find
\begin{eqnarray}
&& \langle 1S_{3/2};F_z^\prime| \delta V ({\bf r}) |
1P_{3/2};F_z\rangle = \frac{v_0}{\sqrt{15}}
\label{eq:mixel1} \\
&& \hspace*{2cm} \times \int dr \, r^2 \, \Bigl[ R_0 R_1 +
\frac{4}{5}R_2 R_1 + \frac{3}{5}R_2 R_3 \Bigr] \nonumber
\end{eqnarray}
for $(F_z^\prime,F_z)=\pm (3/2,1/2)$ or $\pm(1/2,3/2)$, and
\begin{eqnarray}
&& \langle 1S_{3/2};F_z^\prime| \delta V ({\bf r}) |
1P_{3/2};F_z\rangle = \frac{v_0}{\sqrt{5}}
\label{eq:mixel2} \\
&& \hspace*{2cm} \times \int dr \, r^2 \, \Bigl[ \frac{2}{3} R_0
R_1 + \frac{8}{15}R_2 R_1 + \frac{2}{5}R_2 R_3 \Bigr] \nonumber
\end{eqnarray}
for $(F_z^\prime,F_z)=\pm ( 1/2, - 1/2)$. The diagonal matrix
elements $\langle 1S_{3/2};F_z| \delta V ({\bf r}) |1S_{3/2};F_z
\rangle = \pi v_0/4$ and $\langle 1P_{3/2};F_z| \delta V ({\bf r})
|1P_{3/2};F_z \rangle \simeq \pi v_0 (9 + 2
\delta_{|F_z|,1/2})/40$ are evaluated analogously.

The energy eigenstates are calculated by diagonalization
$\hat{H}_{\rm vb} + \hat{H}_{\rm an} + \delta V({\bf r})$ in the
eight-dimensional space spanned by the $1S_{3/2}$ and $1P_{3/2}$
multiplets. Mixing of $2S_{3/2}$ and $3S_{3/2}$ with $1P_{3/2}$ is
neglected because of the large energy difference
[Fig.~\ref{Fig1}(f)]. Because $\hat{H}_{\rm vb}$, $\hat{H}_{\rm
an}$, and $\delta V({\bf r})$ are even under reflection at the
$x$-$y$-plane, the eigensystem consists of four doublets. The
eigenstates
\begin{equation} |\Psi_{v,i}^{(SP)} \rangle =
\sum_{F_z=-3/2}^{3/2} \left( \alpha_{i,F_z} |1S_{3/2};F_z\rangle + \beta_{i,F_z} |1P_{3/2};F_z\rangle \right)
\label{eq:mixstates}
\end{equation}
have components  with $s$- and $p$-type envelope functions with
expansion coefficients $\alpha_{i,F_z}$ and $\beta_{i,F_z}$,
respectively. The corresponding energy eigenvalue is denoted by
$E_{v,i}$.

The mixing of $1S_{3/2}$ and $1P_{3/2}$ multiplets renders
transitions from all eight states $|\Psi_{v,i}^{(SP)} \rangle$ to
$1S_e$ dipole-allowed [shown schematically in Fig.~\ref{Fig3}(b)].
Defining the energy of the $1S_e$-$\Psi_{v,i}^{(SP)}$ exciton by
$E_{X_i}$, for QDQWs with symmetry axes oriented along the laser
beam, the spectral representation for $\theta_F(E)$ reads
\begin{eqnarray}
\theta_F (E) & \simeq & C^\prime E I_1 \sum_{i}
\left(|\alpha_{i,-3/2}|^2 - \frac{|\alpha_{i,1/2}|^2}{3}\right)
\label{eq:frot4} \\ && \hspace*{1cm} \times
\frac{E-E_{X_i}}{(E-E_{X_i})^2 + \Gamma^2},  \nonumber \\
&& +  C^\prime E \sum_{n=2,3} I_n \Bigl[
\frac{E-(E_{X,n}+\tilde{\Delta}_{n,3/2})}{[E-(E_{X,n}+\tilde{\Delta}_{n,3/2})]^2 + \Gamma^2}  \nonumber  \\ &&
\hspace*{0.5cm} - \frac{1}{3} \frac{E-(E_{X,n}+\tilde{\Delta}_{n,1/2})}{[E-(E_{X,n}+\tilde{\Delta}_{n,1/2})]^2 +
\Gamma^2}\Bigr], \nonumber
\end{eqnarray}
where the sum $i$ extends over the states $|\Psi_{v,i}^{(SP)} \rangle$. For the $2S_{3/2}$ and $3S_{3/2}$
valence band multiplets [third and fourth line of Eq.~(\ref{eq:frot4})], $\tilde{\Delta}_{n,F_z} =
\Delta_{n,F_z} + \langle nS_{3/2};F_z|\delta V({\bf r})|nS_{3/2};F_z \rangle$ takes into account the energy
shifts effected by the lattice anisotropy $\hat{H}_{\rm an}$ and the shape anisotropy $\delta V({\bf r})$ to
first order. For $v_0=0$, this expression correctly simplifies to the one obtained for the spherical system.

In experiment, the QDQWs are randomly oriented relative to the
laser beams. For QDQWs with symmetry axes rotated by $\psi$
relative to the laser beam, $\theta_F(E)$ is calculated similarly
to Sec.~\ref{sec:diel} from the matrix elements listed in
Eq.~(\ref{eq:tilt-trel}). The experimentally more relevant
expression for an ensemble of randomly oriented QDQWs is found by
averaging over $\psi$. For $E$ close to the absorption edge, the
ensemble average $\overline{\theta}_F(E)$ is
\begin{eqnarray}
&& \overline{\theta}_F (E) \simeq  \frac{C^\prime E I_1}{36}
\sum_{i} \bigl[15 |\alpha_{i,-3/2}|^2 + |\alpha_{i,-1/2}|^2
\hspace*{0.5cm} \label{eq:frot5}
\\ &&  \hspace*{0.5cm}- 5
|\alpha_{i,1/2}|^2  - 3 |\alpha_{i,3/2}|^2 + 4 \sqrt{3}
(\alpha_{i,3/2}\alpha_{i,1/2} \nonumber \\ && \hspace*{0.5cm}-
\alpha_{i,-3/2}\alpha_{i,-1/2}) \bigr]
\frac{E-E_{X_i}}{(E-E_{X_i})^2 + \Gamma^2}, \nonumber
\end{eqnarray}
where $\alpha_{i,F_z}$ are the expansion coefficients defined in Eq.~(\ref{eq:mixstates}). Transitions involving
$2S_{3/2}$ and $3S_{3/2}$ are modified in an analogous way, but the expressions are omitted here for brevity.
Note that the relative spectral weight of the individual terms changes compared to an oriented sample
[Eq.~(\ref{eq:frot4})].

\begin{figure}
\centerline{\mbox{\includegraphics[width=8.cm]{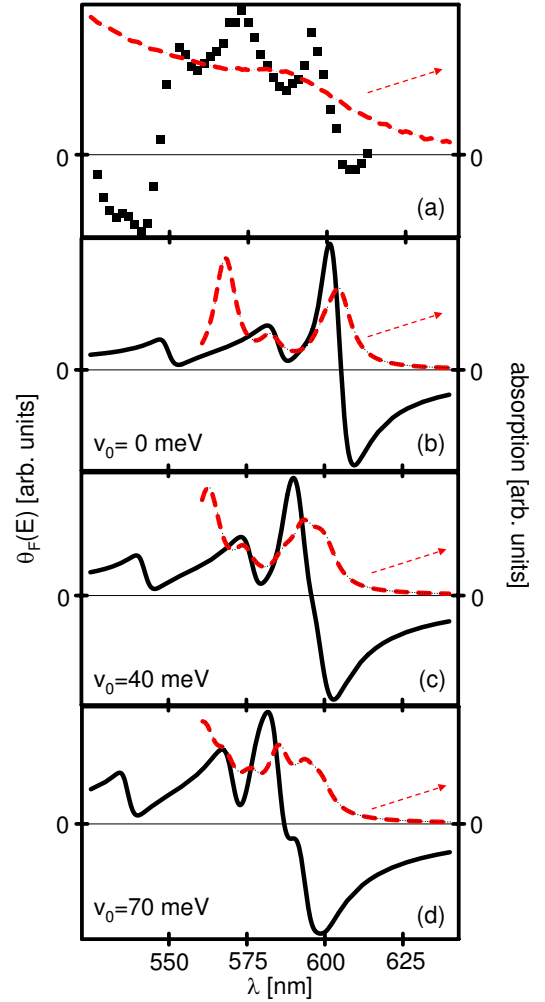}}}
\caption{(color online). (a) Experimental data for the amplitude
of the TRFR signal, $\theta_F(E)$, measured at $T=294 \, {\rm K}$,
and the optical absorption as a function of probe energy. (b)--(d)
Calculated amplitude of the TRFR signal, $\theta_F(E)$, (solid
line) and optical absorption (dashed line) for different strength
of an anisotropy potential, $v_0 = 0, 40, 70 \, {\rm meV}$,
respectively. The calculation is restricted to the $1P_{3/2}$,
$1S_{3/2}$, $2S_{3/2}$, and $3S_{3/2}$ valence band multiplets
which dominate the optical response close to the absorption edge.
For $\lambda < 560 \, {\rm nm}$, transitions from higher valence
band multiplets become important for the absorption spectrum and
the restriction to only four valence band multiplets is no longer
valid. Because $T=294 \, {\rm K}$ is small compared to the
conduction band level splitting, all electron spins are assumed to
occupy $1S_e$ in the calculation of $\theta_F(E)$.}\label{Fig4}
\end{figure}

We next discuss how the TRFR signal amplitude, $\theta_F(E)$, changes with increasing asymmetry potential $v_0$.
Figure~\ref{Fig4}(a) shows the experimental data for $\theta_F(E)$ (symbols) and the absorption spectrum (dashed
line) for the sample with $n_{\rm CdSe} = 3$.~\cite{berezovsky:04} Figures~\ref{Fig4}(b)--(d) show the
calculated TRFR signal amplitudes (solid lines) for (b) $v_0 = 0 \, {\rm meV}$,  (c) $v_0 = 40 \, {\rm meV}$,
and  (d) $v_0 = 70 \, {\rm meV}$, respectively. The TRFR signal for random QDQW orientation is calculated from
Eq.~(\ref{eq:frot5}) taking into account transitions from the $1S_{3/2}$, $2S_{3/2}$, $3S_{3/2}$, and $1P_{3/2}$
valence band multiplets. The level broadening $\Gamma=15\,{\rm meV}$ was chosen to be comparable to the smallest
linewidth in the experimental data. For the system with spherical symmetry, $v_0 = 0 \, {\rm meV}$, the spectral
weight of the resonance with the smallest energy is significantly larger than that of the two resonances at
higher energies, in contrast to the experimental data. With increasing $v_0$, the spectral weight is
redistributed from the $1S_{3/2}-1S_e$ transition to resonances involving other valence band multiplets, such
that the spectral weight of the resonance with lowest energy approaches that of the higher-energy resonances
[Figs.~\ref{Fig4}(c) and (d)]. The corresponding theoretical curves are in better qualitative agreement with the
experimental results than the TRFR signal calculated for a spherically symmetric QDQW.

The single-particle level spectrum and transition matrix elements
calculated above also allow one to calculate the absorption
spectrum,
\begin{equation}
{\rm abs}(E) \propto \sum_{|\Phi_c\rangle; |\Phi_v\rangle} \left|
\langle\Phi_c|\hat{p}_x +  i \hat{p}_y |\Phi_v \rangle \right|^2
\frac{\Gamma}{(E-E_{X})^2 + \Gamma^2}. \label{eq:abs}
\end{equation}
The sum extends over all conduction and valence band states and
$E_X$ denotes the energy of the corresponding exciton transition.
Close to the absorption edge, transitions from $|\Psi_{v,i}^{(SP)}
\rangle$ and the $2S_{3/2}$ and $3S_{3/2}$ multiplets to $1S_e$
and $1P_e$, respectively, are dominant. The larger number of
optically allowed transitions in the system with broken spherical
symmetry [Fig.~\ref{Fig3}(b)] may also explain the absence of
distinct peaks in the absorption spectrum, because transitions
from $p$-type valence band states to $1P_e$ spectrally overlap
with transitions from $s$-type valence band states to $1S_e$. In
Figs.~\ref{Fig4}(b)--(d), we show the evolution of the absorption
spectrum with increasing asymmetry of the QDQW (dashed lines)
calculated using the same parameters as for the TRFR signal. While
the TRFR signal is dominated by transitions involving $nS_{3/2}$
multiplets (Sec.~\ref{sec:diel}), the absorption spectrum also
involves transitions from excited valence band multiplets such as
$1P_{1/2}$ and $2P_{3/2}$ which are neglected here. For the
absorption spectrum, our calculations restricted to the
$1S_{3/2}$, $2S_{3/2}$,  $3S_{3/2}$, and $1P_{3/2}$ multiplets
yield valid results only within $\sim 0.15 \,{\rm eV}$ of the
absorption edge. For a QDQW with spherical symmetry, pronounced
resonances are predicted to appear in the absorption spectrum
close to the absorption edge [Fig.~\ref{Fig4}(b)]. For increasing
$v_0$, the redistribution of spectral weight effected by mixing of
different valence band multiplets leads to a broadening of these
resonances [Fig.~\ref{Fig4}(c)] which, ultimately, evolve into a
step-like feature comparable to the experimental data
[Fig.~\ref{Fig4}(d)].

However, for the small transition linewidth $\Gamma=15\,{\rm meV}$
chosen in Fig.~\ref{Fig4}(d), some resonance features can still be
resolved in the theoretical absorption spectrum. One possible
explanation for the discrepancy with experimental data is that the
linewidth of exciton transitions to $1P_e$ is larger than for
$1S_e$ because of orbital relaxation from $1P_e$ to $1S_e$. We
also point out that, for $v_0 \sim 70 \, {\rm meV}$, the
restriction of our analysis to only four valence band multiplets
($1S_{3/2}$, $2S_{3/2}$, $3S_{3/2}$, and $1P_{3/2}$) can no longer
be justified rigorously. A calculation taking into account all
multiplets in Fig.~\ref{Fig1}(f) would be required to yield
rigorous results for the absorption spectrum in this regime of
strong symmetry breaking.

From our analysis of a simple model for broken spherical symmetry, we conclude that mixing of different valence
band multiplets redistributes the spectral weights in the Lehmann representation of $\theta_F(E)$ and accounts
for a large number of resonances with comparable spectral weight (Fig.~\ref{Fig4}). Broken spherical symmetry is
also consistent with a large Stokes shift and a featureless increase of the absorption
spectrum.~\cite{berezovsky:04}

\section{Summary and Discussion}
\label{sec:disc}

Motivated by recent experiments,~\cite{berezovsky:04} we have
calculated the conduction and valence band states of CdS/CdSe/CdS
QDQWs using ${\bf k}\cdot {\bf p}$ theory. The single-particle
spectrum allowed us to evaluate the amplitude of the TRFR signal
as a function of the probe energy, $\theta_F(E)$, for samples with
well defined and random orientation of symmetry axes,
respectively. For spherical QDQWs, $\theta_F(E)$ exhibits a pair
of resonances for every $nS_{3/2}$ valence band multiplet
[Fig.~\ref{Fig2}(b)], but does not correctly reproduce the spacing
and spectral weight of the experimentally observed resonances. By
contrast, a simple model with broken spherical symmetry reproduces
the experimental data at least qualitatively.

For both the TRFR signal and the absorption spectrum, the model with broken spherical symmetry yields better
agreement with experimental data than the calculations for a spherical QDQW (Fig.~\ref{Fig4}). However, the
theoretical curve does not reproduce all experimental features and large values of $v_0$ are required in order
to explain the experimental data. We, hence, re-emphasize that the discussion in Sec.~\ref{sec:mixing} is only a
qualitative analysis of TRFR for broken spherical symmetry. Additional theoretical work is required to improve
the quantitative understanding of the energy level scheme of CdS/CdSe/CdS QDQWs and of the FR angle as a
function of probe energy.

In particular, more realistic microscopic models for the variation
of the QW width should be considered. Eight-band ${\bf k}\cdot
{\bf p}$ calculations would allow one to quantify whether mixing
of light and heavy hole states with the conduction and split-off
band also reduces the spectral weight of the $1S_{3/2}-1S_e$
exciton transition. If the full eight-band model gives results
comparable to our description, this would further corroborate that
spherical symmetry breaking must be taken into account to
understand the level spectrum of CdS/CdSe/CdS
QDQWs.~\cite{pokatilov:01} Tight-binding calculations of the
energy spectrum~\cite{perez-conde:02,xie:02,bryant:03} and
$\theta_F(E)$ would allow one to explicitly include the atomistic
structure of interfaces and to compare our calculations with a
different theoretical framework.

\acknowledgements This work was supported by ONR and DARPA. We
acknowledge helpful discussion with J. Berezovsky, O. Gywat, W.
Lau, and M. Ouyang.

\end{document}